\documentclass[final,5p,times,twocolumn]{elsarticle} 

\usepackage{lineno,hyperref}
\modulolinenumbers[50]

\journal{Elsevier}

\begin{document}

\begin{frontmatter}

\title{The energy calibration system for CANDLES using (n,$\gamma$) reaction}

\author[address1]{T. Iida\corref{mycorrespondingauthor}}
\address[address1]{Faculty of Pure and Applied Sciences, University of Tsukuba,  Tsukuba, Ibaraki, 305-8571, Japan}
\cortext[mycorrespondingauthor]{Corresponding author}
\ead{tiida@hep.px.tsukuba.ac.jp}

\author[address2]{K. Mizukoshi}
\author[address2]{T. Ohata}
\author[address2]{T. Uehara}
\author[address2]{T. Batpurev}
\author[address3]{W. M. Chan}
\author[address5]{K. Fushimi}
\author[address6]{R. Hazama}
%\author[address2]{K. Ichimura}
\author[address2]{M. Ishikawa}
\author[address2]{H. Kakubata}
\author[address3]{K. Kanagawa}
\author[address2]{S. Katagiri}
\author[address2]{B. T. Khai}
\author[address3]{T. Kishimoto}
\author[address2]{X. Li}
\author[address2]{T. Maeda}
\author[address2]{K. Matsuoka}
\author[address4]{K. Morishita}
\author[address2,address8]{M. Moser}
\author[address4]{K. Nakajima}
\author[address3]{M. Nomachi}
\author[address4]{I. Ogawa}
\author[address2]{M. Shokati}
\author[address7]{K. Suzuki}
\author[address3]{Y. Takemoto}
\author[address3]{Y. Takihira}
\author[address4]{Y. Tamagawa}
\author[address2]{K. Tetsuno}
\author[address3]{V. T. T. Trang}
\author[address3]{S. Umehara}
\author[address2]{S. Yoshida}

\address[address2]{Graduate School of Science, Osaka University, Toyonaka, Osaka 560-0043, Japan}
\address[address3]{Research Center for Nuclear Physics (RCNP), Osaka University, Ibaraki, Osaka 567-0047, Japan}
\address[address4]{Graduate School of Engineering, University of Fukui, Fukui, 910-8507, Japan}
\address[address5]{Department of Physics, Tokushima University, Tokushima 770-8506, Japan}
\address[address6]{Department of Environmental Science and Technology, Osaka Sangyo University, Daito, Osaka 574-8530, Japan}
\address[address7]{The Wakasa-wan Energy Research Center, Tsuruga, Fukui 914-0192, Japan}
\address[address8]{Erlangen Centre for Astroparticle Physics, Friedrich-Alexander-Universit at Erlangen-Nurnberg, 91058 Erlangen, Germany}

\begin{abstract}
CAlcium fluoride for the study of Neutrinos and Dark matters by Low-energy Spectrometer (CANDLES) searches for neutrino-less double beta decay of $^{48}$Ca using a CaF$_2$ scintillator array. 
A high Q-value of $^{48}$Ca at 4,272 keV enabled us to achieve very low background condition, however, at the same it causes difficulties in calibrating the detector's Q-value region because of the absence of a standard high-energy $\gamma$-ray source. Therefore, we have developed a novel calibration system based on $\gamma$-ray emission by neutron capture on $^{28}$Si, $^{56}$Fe and $^{58}$Ni nuclei. In the paper, we report the development of the new calibration system as well as the results of energy calibration in CANDLES up to 9 MeV.

\end{abstract}

\begin{keyword}
\sep double beta decay \sep $^{48}$Ca \sep energy calibration \sep (n,$\gamma$) reaction \sep calcium fluoride
\end{keyword}

\end{frontmatter}

\linenumbers

\section{Introduction}
Because the discovery of neutrino oscillations has revealed the existence of a neutrino mass \cite{NeutrinoOscillation,NeutrinoOscillation2}, the search for a non-neutrino emitting double beta decay ($0\nu\beta\beta$) has become one of the preeminent topics in modern physics.
Double beta decay, which emits two neutrinos ($2\nu\beta\beta$), has been observed by several experiments. However, $0\nu\beta\beta$ has to date remained unobserved.
When observing $0\nu\beta\beta$, the neutrino is noted as a Majorana particle and the lepton number conservation is broken. Accordingly, the asymmetry of matter and antimatter in the present universe can theoretically be explained by the leptogenesis scenario \cite{Leptogenesis}.
As a Majorana particle, the neutrino emitted by a neutron via beta decay can be absorbed by another neutron in the same nucleus. The $0\nu\beta\beta$ thus produces a mono-energetic signature at the end of the $2\nu\beta\beta$ energy spectrum. The rate of $0\nu\beta\beta$ increases alongside the square of the effective neutrino mass; therefore, its measurement can provide information on absolute neutrino mass scale.
Since $0\nu\beta\beta$ is an extremely rare signal, and its half-life is longer than 10$^{26}$ year \cite{0nbb,0nbb2}, low radioactive contamination and good energy resolution is required to separate a mono-energetic signal from background noise.

CAlcium fluoride for the study of Neutrinos and Dark matters by Low energy Spectrometer (CANDLES) is a $^{48}$Ca double beta decay experiment using a CaF$_2$ scintillator and a photomultiplier tube (PMT) \cite{CANDLES}. The CANDLES III detector was constructed at the Kamioka Underground Observatory in Japan and is currently collecting experimental data.
The Q-value of $^{48}$Ca (i.e., the sum of the kinetic energies of the two electrons emitted at $0\nu\beta\beta$) is 4,272 keV. It is the largest among the candidate nuclei for double beta decay. Taking advantage of this, CANDLES aims to measure $0\nu\beta\beta$ in an ultra-low background environment \cite{ngamma1}. The high Q-value of $^{48}$Ca makes it difficult, however, to precisely calibrate the detector due to the absence of a high-energy standard gamma ray source. 
To date, we have calibrated the detector using a 1,836 keV $\gamma$-ray obtained from a $^{88}$Y source. This is relatively high energy among commercially available radioactive $\gamma$-ray sources. In addition, we also employed a 2,615-keV $\gamma$-ray from $^{208}$Tl, the source of which was radioactive contamination in the detector components.
A higher energy calibration source is required because precise energy calibration near the Q-value region is very important for the identification of $0\nu\beta\beta$. We would like to report construction of a energy calibration system above 3 MeV using gamma rays from neutron capture reactions on Si, Fe and Ni. In this paper, we also report the energy calibration of the CANDLES detector using the above system.

\section{Detector Overview}
\begin{figure*}[htbp]
 \begin{center}
 \includegraphics[width=14cm]{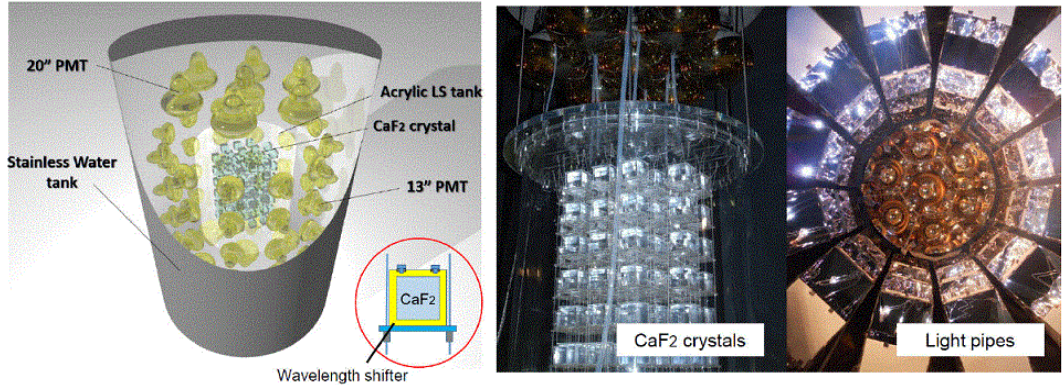}
\caption{The figure on the left presents a schematic view of the CANDLES III detector. The stainless steel tank has a 3-m diameter $\times$ and a 4-m height. The acrylic tank is 1.4 m in diameter $\times$ and has a height of 1.4 m. The size of the CaF$_2$ crystal is 10$^3$ cm$^3$ cubic. Light pipes have been omitted from the figure for better visibility. Two images on the right show the CANDLES CaF$_2$ crystal array and light pipes seen from above the tank.}
\label{fig:CANDLES}
 \end{center}
\end{figure*}

The CANDLES III detector (Figure \ref{fig:CANDLES}) operates with 305-kg CaF$_2$ crystals in the Kamioka Underground Observatory \cite{TAUP}. The detector comprises 96 pure calcium fluoride (CaF$_2$) crystals sized 10$^3$ cm$^3$ immersed in a liquid scintillator (LS) as a 4$\pi$ active shield. 
The crystals were divided into six layers in a vertical (Z) direction, and each layer included 16 crystals.
The LS has a much shorter decay time constant ($\sim$10 ns) compared with CaF$_2$ ($\sim$1 $\mu$s). Events of multiple energy deposition could thus be rejected by pulse shape analysis.
Scintillation lights from CaF$_2$ and the LS were observed by 62 PMTs with light-collecting pipes mounted on a 30 m$^3$ stainless steel water tank.
Since CaF$_2$ crystals are generally used in camera lenses and other commercial applications, comparative crystals with high transparency and high purity have been developed and are available. 
Because of the scintillation light's high transparency and attenuation length longer than 10 m, it was collected by PMTs without attenuation and accurate energy information was obtained.
A scintillation light wavelength of the CaF$_2$ crystals was shifted from 280 to 420 nm by a wavelength shifter (Figure \ref{fig:CANDLES}) to effectively collect photons with PMTs \cite{WLS} .
The CANDLES experiment in this work adopted a data acquisition (DAQ) system using the SpaceWire protocol and the DAQ-Middleware framework. 
To achieve an effective trigger for the event using a long decay time constant of CaF$_2$, dual gate trigger logic was developed. Following the trigger, 8-bit 500 MHz flash analog-to-digital converters (FADCs) opened a 8,960-ns window and read out all PMT waveforms.
Details of the CANDLES DAQ system are summarized in \cite{DAQ,Trig}.

The entire experiment room was cooled to roughly 2 $^{\circ}$C by a cooling system. The detector temperature was retained at roughly 4 $^{\circ}$C with a stability degree of $\pm$ 0.2.
Light yield of the CaF$_2$ scintillator is known to vary inversely according to temperature. With a temperature lowered by 1 $^{\circ}$C, CaF$_2$ will yield a roughly 2 \% larger light output. High-purity crystals can achieve very low radioactive impurities, equating to low background radiation. Radioactive impurity concentration for each crystal was distributed between a few to a few tens [$\mu$Bq/kg] in the Th chain.

\section{Current Calibration in CANDLES}
Energy calibration is important in double beta decay experiments because it determines the region where signals are expected but nothing appears usually. 
We calibrated each crystal with the aim of achieving 0.5\% precision that is adequately smaller than the energy resolution at the Q-value region. Since internal or external background events are poor in statistics for the purpose, we set a radiation source in the tank to acquire large statistics within a short period of time.
We selected $\gamma$-ray of $^{88}$Y source (898 keV and 1,836 keV), which is commercially available, for relative gain calibration between crystals, to best meet the energy and statistics requirements.
The $^{88}$Y source with a maximum of 50 kBq intensity was installed in the LS tank and set inside the crystal array. Events in each crystal were selected via position reconstruction, and CaF$_2$ crystal events were selected by pulse shape analysis, which removed LS energy deposition larger than approximately 100 keV at 1,836 keV.
Based on the above procedure, we successfully performed energy calibration for each crystal using a $\gamma$-ray energy peak at 1,836 keV within 0.3 \% statistical precision.

After relative calibration by the $^{88}$Y source, we corrected the absolute energy scale using an external $\gamma$-ray from $^{208}$Tl, the energy of which was 2,615 keV closer to the Q-value of $^{48}$Ca. This $\gamma$-ray was derived from parts of the detector and its surrounding material. Thus, the event rate was not high enough to affect crystal-by-crystal calibration with good precision.
Since we found a small \% layer dependence of energy scale in the 2,615-keV $\gamma$-ray, even after $^{88}$Y calibration, energy scale correction was applied for each layer (i.e., 16 crystals).

\section{New Calibration Method}
\begin{figure}[htbp]
 \begin{center}
 \includegraphics[width=8cm]{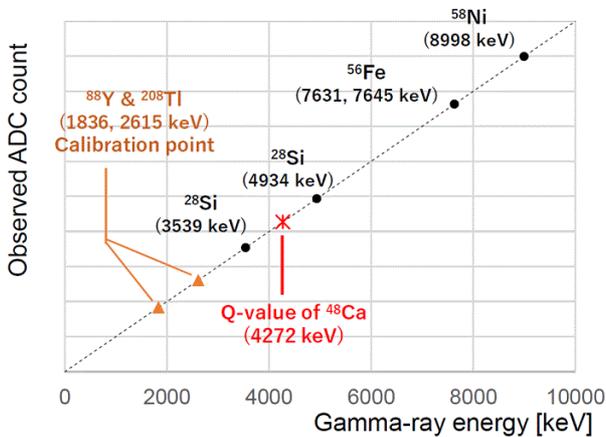}
\caption{Conceptual plot for the CANDLES energy calibration. The cross marker shows the Q-value of $^{48}$Ca; the triangle indicates the current calibration point at 1,836 keV for $^{88}$Y and 2,615 keV for $^{208}$Tl. The black circle shows the new calibration point added by this paper.}
\label{fig:intro}
 \end{center}
\end{figure}

Since Q-value was significantly higher than the current calibration point at 2,615 keV, we had to check the energy scale linearity at 4,272 keV. We developed a high-energy $\gamma$-ray calibration source by means of (n,$\gamma$) reaction.
In general, several MeV $\gamma$-rays were emitted by neutron capture reaction on various nuclei.
The $^{28}$Si (n,$\gamma$) reaction was to some extent unique. The excited state of $^{29}$Si following neutron capture emitted a 3,539-keV $\gamma$-ray by decaying to an excited state at 4,934 keV, which in turn emitted a 4,934-keV $\gamma$-ray \cite{ngamma2,ngamma3}. 
Since the energies of these $\gamma$-rays were close to the Q-value, they were able to provide an ideal energy calibration for CANDLES. However, the small cross section of 0.17 barn for thermal neutron capture reaction required an effective neutron moderation system. This will be described later in this paper.
Other neutron-capture $\gamma$-rays used in this system are summarized in table \ref{table:ngamma}. 
With the above $\gamma$-rays, the energy scale at Q-value (4,272 keV) could be obtained by interpolation. To date, however, we have relied on extrapolation (see Figure \ref{fig:intro}).

In this chapter, we describe the development of a new calibration system using the (n,$\gamma$) reaction of $^{28}$Si, $^{56}$Fe, and $^{58}$Ni to affect calibration in the energy region of 3 MeV and higher.

\begin{table}[hbtp]
  \caption{The $\gamma$-rays from the thermal neutron capture used in this calibration system. The neutron capture cross section emitted $\gamma$-ray energy and a $\gamma$ decay branching ratio from an excited state (refer to database \cite{iaea}.)}
  \label{table:ngamma}
  \centering
  \begin{tabular}{cccc}
    \hline
      & Neutron cross section &  $\gamma$-ray energy  & Branching ratio \\
    \hline \hline
    $^{28}$Si  & 0.177 barn  & 3,539 keV & 67.2 \% \\
& & 4,934 keV & 63.3 \% \\
    $^{56}$Fe  & 2.59 barn & 7,631 keV & 25.2 \% \\
      &  & 7,645 keV & 21.2 \% \\
    $^{58}$Ni  & 4.50 barn & 8,998 keV & 33.1 \%\\
    \hline
  \end{tabular}
\end{table}

\subsection{Development of Silicon and Nickel blocks}
Considering expandability and portability, we developed polyethylene blocks including metallic silicon (Si) and nickel oxide (NiO), referred to as a Si or Ni block, respectively. Each block was 20 cm $\times$ 10 cm $\times$ 5 cm. An image of a Si block is shown at the bottom right in Figure \ref{fig:setup}. 
Powdered Si or NiO was mixed with polyethylene and solidified using an epoxy adhesive. 
Since polyethylene and epoxy adhesive include large amounts of hydrogen atoms, fast neutrons are efficiently thermalized and captured on Si or Ni nuclei inside blocks.
The composition of Si and Ni blocks are summarized in Table \ref{table:block}.
In total, thirty-six and eight blocks were made for Si and Ni, respectively, to obtain a sufficient amount of (n,$\gamma$) events under CANDLES detection efficiency.

\begin{table}[hbtp]
  \caption{Composition of Silicon and Nickel blocks}
  \label{table:block}
  \centering
  \begin{tabular}{ccc}
    \hline
      & Component material &  Weight ratio  \\
    \hline \hline
    Si block & Metallic silicon (Si) & 63.5 \% \\
     & Polyethylene & 16.5 \% \\
     & Epoxy adhesive & 20.0 \% \\
   \hline
    Ni block & Nickel oxide (NiO) & 22.5 \% \\
     & Polyethylene   & 42.5 \% \\
     & Epoxy adhesive & 35.0 \% \\
    \hline
  \end{tabular}
\end{table}

\subsection{Calibration Setup on CANDLES}
\begin{figure}[htbp]
 \begin{center}
 \includegraphics[width = 8 cm]{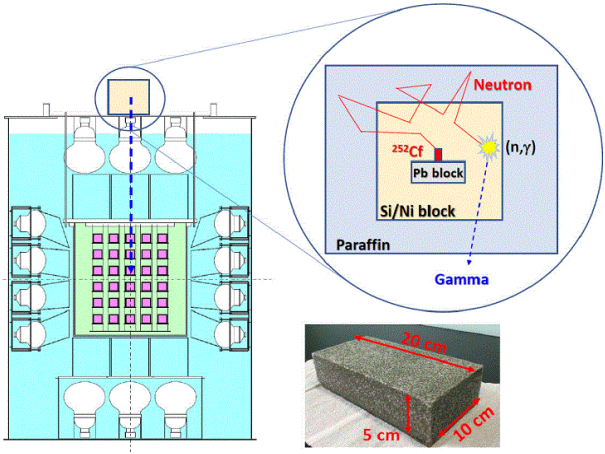}
\caption{Schematic view of the calibration system. Si and Ni blocks were set on top of the CANDLES detector. Neutrons emitted from $^{252}$Cf were moderated by collision on hydrogen atoms and finally captured by $^{28}$Si or $^{58}$Ni nuclei, followed by $\gamma$-ray emission. The bottom right image shows a Si block sized 20 cm $\times$ 10 cm $\times$ 5 cm.}
 \label{fig:setup}
 \end{center}
\end{figure}

Calibration data were taken with the blocks set on top of the CANDLES detector (see Figure \ref{fig:setup}).
The $^{252}$Cf neutron source, the decay rate of which was roughly 200 kBq, was placed at the center of the blocks.
The decay of $^{252}$Cf included 3\% spontaneous fission and emitted an average of 3.8 neutrons with a mean energy of approximately 2 MeV.
Fast neutrons were moderated by repeated collision with hydrogen atoms inside the blocks or in the surrounding paraffin blocks.
The $\gamma$-ray emitted by moderated neutron capture on Si or Ni nuclei was detected by the CANDLES detector and used for detector calibration.
Since $^{252}$Cf emitted not only neutrons but also many $\gamma$-rays, a 5-cm Pb shield was placed in the CANDLES direction in order to reduce such background $\gamma$-rays.

We conducted three types of calibration runs with different setups, as well as a background run.
The Si run used 36 Si blocks and the Ni run employed eight Ni blocks; blocks corresponded to 34 kg Si and 2.3 kg Ni, respectively.
In the $^{252}$Cf run, the $^{252}$Cf source was placed on the stainless steel tank directly in order to generate the $\gamma$-ray from the $^{56}$Fe neutron capture contained in the stainless steel tank.
The conditions of three calibration runs and a background run are summarized in table \ref{table:run}.

\begin{table}[hbtp]
  \caption{Summary of the Calibration Runs for Each Setup}
  \label{table:run}
  \centering
  \begin{tabular}{ccc}
    \hline
   Run & Setup & Duration \\
    \hline \hline
    Si run & Si block (34 kg of Si) & 50.7 hours \\
    Ni run & Ni block (2.3 kg of Ni) & 12.1 hours \\
    $^{252}$Cf run & $^{252}$Cf on tank & 12.0 hours \\
    Background run & No source & 131 days \\
    \hline
  \end{tabular}
\end{table}

\section{Data Analysis}
\begin{figure}[htbp]
 \begin{center}
 \includegraphics[width=8cm]{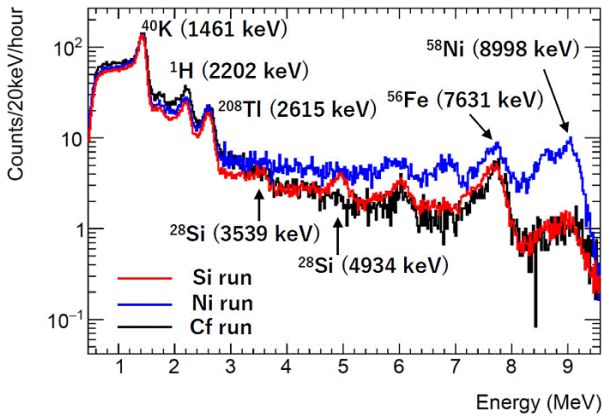}
\caption{Energy spectra for Si run (red), Ni run (blue), and $^{252}$Cf run (black) following LS cut normalized by the live time of each run. Characteristic peaks appear in each run.}
 \label{fig:Spec}
 \end{center}
\end{figure}

Detailed analysis of the obtained data was carried out offline.
Figure \ref{fig:Spec} shows the energy spectra in three different setups using observed energy calibrated by $^{88}$Y and $^{208}$Tl.
Event selection (LS cut) removed more than 99 \% of events with LS energy deposition of 100 keV at the visible energy of 1,836 keV. This was applied to observe clear peaks through a reduction of Compton scattering events.
The $\gamma$-rays' energy peaks from $^{28}$Si, $^{56}$Fe, and $^{58}$Ni were apparent in the data of the Si, $^{252}$Cf, and Ni runs, respectively.

\begin{figure*}[htbp]
 \begin{center}
 \includegraphics[width = 13.5 cm]{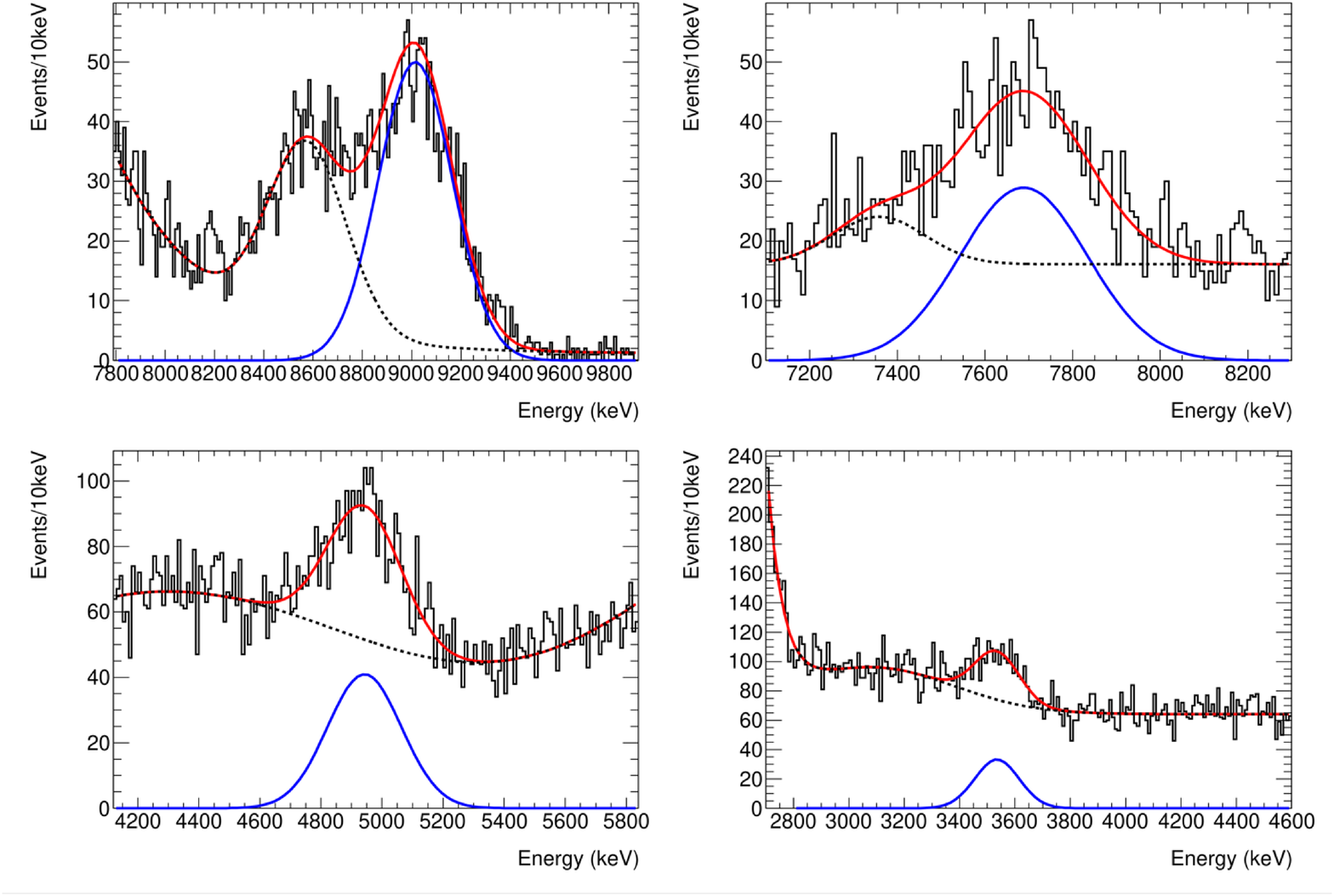}
\caption{Fitting results for 8,998 keV of $^{58}$Ni (top left), 7,631 $+$ 7,645 keV of $^{56}$Fe (top right), 4,934 keV of $^{28}$Si (bottom left), and 3,539 keV of $^{28}$Si (bottom right), respectively. The red line shows the fitting results, the black dashed line shows the background, and the blue line represents the signal}.
 \label{fig:Fit}
 \end{center}
\end{figure*}

In order to extract a signal from the obtained spectrum, fitting was performed under the assumption that the background components comprised Gaussian, exponential, polynomial, and constant elements.
Figure \ref{fig:Fit} show the fitting results of each $\gamma$-ray energy peak. The two upper images show $^{58}$Ni and $^{56}$Fe; the bottom left represents 4,934 keV of $^{28}$Si, and the bottom right that of 3,539 keV of $^{28}$Si.
In the case of Ni, an exponential, a constant, and a Gaussian element were assumed as background for the 8,533 keV $\gamma$-ray of $^{58}$Ni in order to produce the 8,998-keV peak of $^{58}$Ni. In this figure, 8,533 keV of $^{58}$Ni is visible; however, it was not used for calibration because this energy region may have been affected by the Compton edge of the 8,998-keV $\gamma$-ray.
The $^{56}$Fe emitted two very close energy $\gamma$-rays with energies of 7,631 and 7,645 keV and branching ratios of 25.2 \% and 21.2 \%, respectively. 
Although data included these two $\gamma$-rays, their energy difference was much smaller than the energy resolution of CANDLES ($\sigma \sim$ 140 @ 7,631 keV). These two peaks were thus treated as one Gaussian distribution with a mean of 7,637 keV, the weighted average of two $\gamma$-rays. In addition, the 7,279 keV $\gamma$-ray of $^{56}$Fe and a constant component were assumed as background during fitting.
In the two Si analyses, the background was unknown. The shape of the background was thus assumed by an appropriate combination of Gaussian, exponential, polynomial, and constant elements.
We attempted several different combinations and confirmed that the results varied within a statistical error at most.
Due to the relatively small cross section of $^{28}$Si, it was anticipated that contamination by higher energy $\gamma$-ray events (which partially deposited energy into CaF$_2$ by Compton scattering) could render analysis difficult.
However, the obtained results from $^{28}$Si analyses were reasonable.

\begin{figure}[htbp]
 \begin{center}
 \includegraphics[width=8cm]{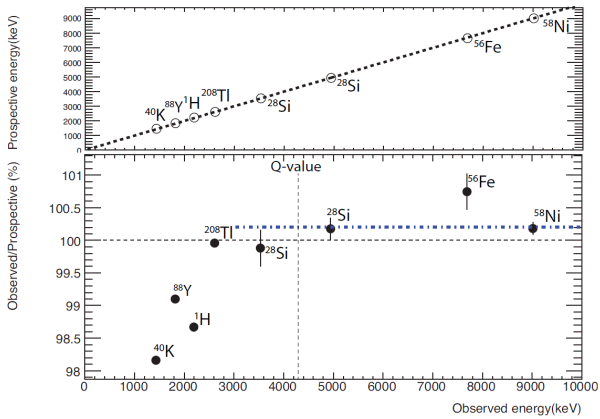}
\caption{The top image shows prospective energy vs observed energy. Each point corresponds to $\gamma$-ray peak energy measured by calibration or background run.
The bottom image shows energy deviation for CANDLES as a function of observed energy. Blue dash-dot line is an average obtained from four new calibration points.  
Good linearity of energy response within 0.3 \% was obtained in the region of interest by this work.
Error bar shows only statistical error.}
 \label{fig:lin}
 \end{center}
\end{figure}

The energy linearity and energy resolution of CANDLES were studied using the results of each peak fitting.
Figure \ref{fig:lin} shows the linearity of energy response as a function of observed $\gamma$-ray energy. The Y-axis of the bottom plot is a ratio of observed energy and prospective energy for each $\gamma$-ray.
Observed energy is peak energy extracted by fitting. Prospective energy refers to the database of the International Atomic Energy Agency \cite{iaea}.
Four additional calibration points above 3 MeV became available following installation of the new calibration system and the energy scale at the Q-value region is now obtained by interpolation.
From this figure, good energy linearity for CANDLES up to 9 MeV was confirmed and energy deviation at the Q-value region was estimated to be less than 0.3 \%.

\begin{figure}[htbp]
 \begin{center}
 \includegraphics[width=8cm]{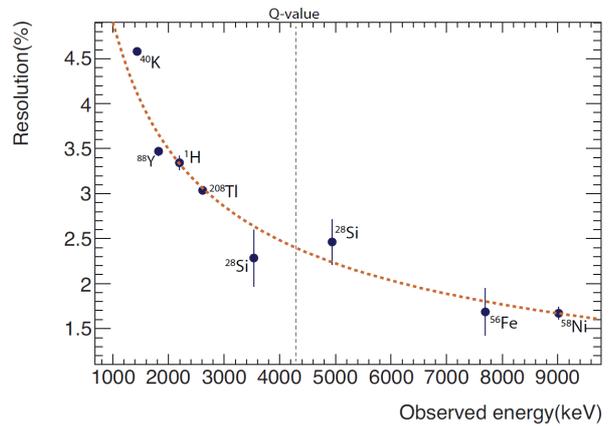}
\caption{Energy resolution ($\sigma$) for CANDLES as a function of observed energy. The circle marker indicates $\gamma$-ray observed in CANDLES by calibration and background run.
Single exponential function indicated good fit with the data up to 9 MeV (dashed line); $^{40}$K near the detector threshold was removed from fitting points. The error bar indicates only statistical error.}
 \label{fig:res}
 \end{center}
\end{figure}

Figure \ref{fig:res} shows the energy resolution for CANDLES as a function of observed energy. Each circle point corresponded to $\gamma$-ray measured in the calibration or background run. The Y-axis of the plot was a fitted $\sigma$ for each $\gamma$-ray peak.
Simple single exponential fitting resulted in roughly $\sigma$ = 2.4 $\pm$ 0.2 \% at 4,272 keV.

\section{Conclusion and Outlook}
In this paper, we reported a new energy calibration method for CANDLES to search for neutrino-less double beta decay of $^{48}$Ca.
A high Q-value of $^{48}$Ca may be a key to performing very low background $0\nu\beta\beta$ measurements.
To date, an absolute energy scale of CANDLES was calibrated at 2,615 keV, despite a high Q-value of 4,272 keV.
We designed and developed the calibration method by using $\gamma$-rays from (n,$\gamma$) reactions. This enabled us to calibrate energy up to 9 MeV.

Four calibration points at 3,539 keV, 4,934 keV of $^{28}$Si, 7,631 $+$ 7,645 keV of $^{56}$Fe, and 8,998 keV of $^{58}$Ni were made available by this work.
The obtained results provide confirmation of good energy linearity above 3 MeV for the first time in a CANDLES experiment with uncertainty below 0.3 \% at 4,272 keV, which is adequately smaller than the resolution.
Additionally, energy resolution at the Q-value was measured and estimated to be 2.4 $\pm$ 0.2 \% at 4,272 keV.

In future research, the background must be reduced in order to see clear peaks and to limit the number of statistical errors.
Peaks of $^{28}$Si were contaminated by the 7,631-keV $\gamma$-ray of $^{56}$Fe, which partially deposited the energy by Compton scattering in CaF$_2$.
An additional neutron shield on the stainless steel tank in the Si run should effectively reduce such background noise and ensure better energy calibration for the CANDLES experiment.

\section*{Acknowledgement}
This work was supported by the JSPS KAKENHI Grant-in-Aid for Scientific Research (B) 18H01222 and (S) 24224007 as well as Grant-in-Aid for Scientific Research on Innovative Areas 26105513, 16H00870, 26104003. 
The Institute for Cosmic Ray Research (ICRR) hosted and partially supported the CANDLES experiment. 
The authors would like to thank the Kamioka Mining and Smelting Company for their assistance with the mine and the CI kogyo Company for manufacturing the Si and Ni blocks.
Finally, the authors acknowledge Mr. Nobuo Nakatani, a great technician who retired in 2016, who supported our experimental work in the mine.

\section*{References}

\bibliography{mybibfile}

\begin{thebibliography}{10}

\bibitem{NeutrinoOscillation}
Y.~Fukuda et~al.
\newblock Evidence for oscillation of atmospheric neutrinos.
\newblock {\em Phys. Rev. Lett.}, 81:1562, 1998.

\bibitem{NeutrinoOscillation2}
Q.~R.~Ahmad et~al.
\newblock Direct evidence for neutrino flavor transformation from
  neutral-current interactions in the sudbury neutrino observatory.
\newblock {\em Phys. Rev. Lett.}, 89:011301, 2002.

\bibitem{Leptogenesis}
M.~Fukugita and T.~Yanagida.
\newblock Barygenesis without grand unification.
\newblock {\em Phys. Lett. B}, 174:45--47, 1986.

\bibitem{0nbb}
A.~Gando et~al.
\newblock {Search for Majorana Neutrinos Near the Inverted Mass Hierarchy
  Region with KamLAND-Zen}.
\newblock {\em Phys. Rev. Lett.}, 117:082503, 2016.

\bibitem{0nbb2}
M.~Agostini et~al.
\newblock {Improved Limit on Neutrinoless Double-$\beta$ Decay of 76Ge from
  GERDA Phase II}.
\newblock {\em Phys. Rev. Lett.}, 120:132503, 2018.

\bibitem{CANDLES}
T.~Kishimoto et~al.
\newblock {Candles for the study of beta beta decay of Ca-48}.
\newblock {\em 4th workshop on Neutrino Oscillations and their Origin}, page
  338, 2003.

\bibitem{ngamma1}
K.~Nakajima et~al.
\newblock {Background studies of high energy $\gamma$ rays from (n,$\gamma$)
  reactions in the CANDLES experiment}.
\newblock {\em Astropart. Phys.}, 100:54--60, 2018.

\bibitem{TAUP}
T.~Iida et~al.
\newblock {Status and future prospect of 48Ca double beta decay search in
  CANDLES}.
\newblock {\em J. Phys. Conf. Ser.}, 718:062026, 2016.

\bibitem{WLS}
S.~Yoshida et~al.
\newblock {Ultra-violet wavelength shift for undoped CaF$_2$ scintillation
  detector by two phase of liquid scintillator system in CANDLES}.
\newblock {\em Nucl. Instrum. Meth.}, A601:282--293, 2009.

\bibitem{DAQ}
K.~Suzuki et~al.
\newblock {New DAQ System for the CANDLES Experiment}.
\newblock {\em Nuclear Science, IEEE Transactions}, 62(3):1122--1127, 2015.

\bibitem{Trig}
T.~Maeda et~al.
\newblock {The CANDLES Trigger System for the Study of Double Beta Decay of
  48Ca}.
\newblock {\em Nuclear Science, IEEE Transactions}, 62(3):1128--1134, 2015.

\bibitem{ngamma2}
A.~M. F. OP DEN~KAMP A.~M. J.~SPITS and H.~GRUPPELAAR.
\newblock {GAMMA RAYS FROM TItERMAL-NEUTRON CAPTURE IN NATURAL AND 28Si
  ENRICHED SILICON }.
\newblock {\em Nuclear Phys. A}, 145:449--460, 1970.

\bibitem{ngamma3}
G.~B. BEARD and G.~E. THOMAS.
\newblock {GAMMA RAYS FROM THERMAL NEUTRON CAPTURE IN $^{28}$Si, $^{29}$Si AND
  $^{30}$Si}.
\newblock {\em Nuclear Phys. A}, 157:520--528, 1970.

\bibitem{iaea}
{Database for Prompt Gamma-ray Neutron Activation Analysis:} available:
  https://www-nds.iaea.org/pgaa/.

\end{thebibliography}

\end{document}